# ارائه روشی سریع جهت ترکیب الگوریتم های رمزنگاری AES و فشرده سازی LZ4

صابر ملک زاده[1]

1- گروه علوم کامپیوتر، دانشگاه ولیعصر(عج) رفسنجان، رفسنجان، ایران

## خلاصه

از آغاز بشریت تاکنون، همواره در کنار رمزنگاری داده ها و افزایش امنیت آنها، کاهش حجم اطلاعـات و بسـته بنـدی داده ها به گونه‌ای که باعث بهبود سرعت انتقال داده‌ها شود مورد بحث بوده است. به خصوص در سال های اخیر که نیاز به بهبود روش‌های رمزنگاری و فشرده‌سازی به جهت انتقال سریع و آسان اطلاعات بیش از پیش حس می‌شود.
در این مقاله، روشی جدید جهت ترکیب دو الگوریتم فشرده‌سازی LZ4 و الگوریتم رمزنگاری AES جهت رسیدن بـه سرعت مناسب، کاهش حجم داده‌ها، بسته‌بندی سریع و امن کردن داده‌ها ارائه مـی‌شـود. انتخـاب دو الگـوریتم مـذکور بـه جهت ویژگی‌های خاص آنها مناسب با هدف مقاله است. این مقاله به شرح این روش می‌پردازد که بـا مـوازی سـازی عمـل فشرده‌سازی و رمزنگاری به شیوه‌ای خاص بهبودی محسوس در سرعت و امنیت انتقال داده‌ها ایجاد خواهد شد.

واژه های کلیدی: LZ4، AES، فشرده سازی، رمزنگاری، سرعت انتقال.

## 1. مقدمه

در سالهای اخیر با گسترش روش‌های متعدد رمزنگاری داده کار بر روی بهبود هر چه بیشتر این روش‌ها نیز گسـترش یافته است. همچنین جهت افزایش سرعت، بسته بندی، کاهش حجم و انتقال آسان داده‌ها تاکنون الگوریتم هـای فشـرده‌-سازی متنوعی با اهداف مختلف مطرح شده‌اند. در میان الگوریتم‌های مطرح در رمزنگاری داده یکی از سـریع‌تـرین و امـن-تـرین آنها الگوریتم رمزنگاری AES است. AES بـا مقالـه‌ای [1] بـه سـال 2001 توسـط موسسـه ملـی فنـاوری و استانداردهای ایالات متحده آمریکا به عنوان برنده استاندارد پیشرفته رمزنگاری انتخاب و معرفی شـد. ایـن الگـوریتم جـزء الگوریتم‌های رمزنگاری کلید متقارن می‌باشد، از معماری فیستلی برای رمزکردن داده‌ها استفاده می‌کند و به جهت سـرعت و امنیت بالا هم اکنون به عنوان الگوریتم استاندارد در اکثر سیستم های انتقال داده به کار برده می‌شود.
LZ4 یکی از جدیدترین و سریع‌ترین الگوریتم‌های فشرده‌سازی از خانواده الگوریتم‌های فشرده‌سازی بایت‌گرا و بی‌اتفلاف و از وارثان الگوریتم LZ77 ارائه‌شده در مقاله‌ای [2] به سال 1977 می‌باشد. این الگوریتم که کمتر از سه سال از ارائه اولیه آن می‌گذرد، در فوریه سال 2014 نسخه پایدار آن نیز منتشر شد. از مهم ترین ویژگی‌های ایـن الگـوریتم کـاهش مناسـب حجم داده‌ها و در عین حال سرعت بسیار بالا مخصوصا در هنگام بازسازی فایل فشرده است.
در سال‌های اخیر تلاش هایی برای بهبود فشرده سازی‌ها و امن‌کردن آن‌ها صورت گرفته که از مهم ترین آن‌ها می توان بـه استفاده سیستم‌های مدیریت پایگاه داده اشاره نمود. با وجود این در اکثر این سیستم‌ها کاستی‌هایی به چشم می‌خـورد کـه از میان آن‌ها می توان به عدم موازی‌سازی صحیح دو عملیات رمزنگاری و فشرده‌سازی، عدم استفاده از الگوریتم‌های مناسب با هدف و همچنین پیاده‌سازی غیربهینه الگوریتم در آن‌ها اشاره نمود.

در این مقاله به ارائه روشی خاص جهت ترکیب دو الگوریتم AES و LZ4 به صورت موازی پرداخته شده است تا به واسطه کاهش حجم داده‌ها، رمزنگاری آن‌ها و در عین حال موازی سازی این دو روش به گونه‌ای خاص، سرعت انتقال داده‌ها به صورت امن افزایش داده شود.

## 2. کارهای مشابه

فعالیت‌های متعددی در زمینه‌ی بهبود عملکرد و سرعت بخشیدن به انتقال اطلاعات در گذشته صورت پذیرفته است. استفاده از الگوریتم‌های مختلف رمزنگاری برای امن کردن فضای انتقال اطلاعات و استفاده از الگوریتم‌های فشرده‌سازی برای کاهش حجم اطلاعات. شاید مهم‌ترین و مشهورترین الگوریتم رمزنگاری همان AES باشد. با ورود LZ4 به عرصه‌ی فشرده‌سازی اطلاعات و سریع بودن فشرده‌سازی توسط این الگوریتم نیاز برای ترکیب این دو الگوریتم برای بهبود سرعت انتقال اطلاعات به خصوص در فضای وب بیش از پیش حس می‌شود.

## 3. LZ4

در سال های اخیر با گسترش ارتباطات و انتشار داده‌های بزرگ و نیاز به کاهش حجم داده بدون از دست دادن بخش‌هایی از آن به وضوح حس می‌شود. فشرده‌سازی داده با هدف کاهش حجم آن در سال های اخیر گسترش فراوانی داشته‌است و الگوریتم‌های مختلفی برای این کار ارائه شده‌است که اکثر آن‌ها با اهداف خاص توسعه داده شده‌اند. برای مثال هر فرمت دارای الگوریتم بهینه ای برای فشرده سازی بوده‌است. در این میان الگوریتم‌های دیگری نیز بوده اند که برای فشرده سازی تمام انواع فایل ها به صورت عمومی ارائه شده اند که البته باتوجه به گستردگی دامنه، درصد فشرده‌سازی کمتری نسبت به روش‌های فشرده‌سازی فرمت‌های خاص داشته‌اند. LZ4 یکی از همین الگوریتم‌هاست. این الگوریتم بر روی هر دو فاکتور مهم فشرده سازی یعنی درصد فشرده‌سازی و سرعت تمرکز دارد به گونه ای که هم بوسیله روش Arithmetic coding درصد فشرده‌سازی را افزایش داده و هم با کاهش تعداد اعمال محاسباتی و استفاده از واژه نامه، سرعت فشرده‌سازی را تا حد قابل ملاحظه‌ای افزایش می‌دهد.

### 3.1. سرعت الگوریتم LZ4

در اهدافی که برای طرح کردن الگوریتم LZ4 اعلام شده‌است سرعت حرف اول را می‌زند. نتایج به دست آمده از آزمون‌هایی که بر روی این الگوریتم صورت گرفته‌است گویای این مطلب است. در مقایسه با سریع‌ترین روش‌های فشرده‌سازی داده‌ها الگوریتم LZ4 به طور میانگین سرعتی 1.5 برابر سریع‌تر در فشرده‌سازی و 3 برابر سریع‌تر در بازسازی فایل فشرده ارائه می نماید. جدول زیر که در بخش Code وبسایت Google ارائه شده، برپایه پردازنده Core i5-3340M @2.7GHz که تنها با یک پردازه (نخ) اجرا شده است.

```
Name              Ratio     C.speed   D.speed
                            MB/s      MB/s

LZ4 (r101)        2.084     422       1820
LZO 2.06          2.106     414        600
QuickLZ 1.5.1b6   2.237     373        420
Snappy 1.1.0      2.091     323       1070
LZF               2.077     270        570
zlib 1.2.8 -1     2.730      65        280
LZ4 HC (r101)     2.720      25       2080
zlib 1.2.8 -6     3.099      21        300
```

جدول (1) : مقایسه سرعت الگوریتم LZ4 [7]

## 2.3. فشرده سازی

در ابتدا باید گفت هدف اصلی الگوریتم LZ4 افزایش سرعت فشرده سازی داده‌هاست. با این حال تمرکـز بـر روی میزان فشرده‌سازی داده‌ها نیز کم نبوده‌است. در مقایسه با الگوریتم‌ها و نرم‌افزارهای امروزی میزان فشرده‌سازی در الگوریتم LZ4 تا 10 درصد کمتر است. اما با مقایسه سرعت فشرده‌سازی می توان متوجه این نکته شد که به خصـوص در تبـادلات تحت شبکه برای مثال تبادلات اینترنتی، درصد بهبود سرعت در LZ4 (50٪) نسبت به دیگر الگوریتم‌ها بسیار باصرفه‌تر از کاهش میزان فشرده‌سازی (10٪) در آن است. در نتیجه انتخاب این الگوریتم انتخاب مناسبی خواهدبود.

## 3.3. ساختار بلوک‌ها

در مقاله ای [3] که Yann Collet در سال 2013 ارائه‌نمود به بیان گسترده کلیات بلوک‌های فشرده‌سازی در روش LZ4 پرداخت و چگونگی فشرده‌سازی داده‌ها در قالب این بلوک‌ها را شرح داد که به اختصار به شرح آن پرداخته می‌شود.

شکل (1) : بلوک داده ای فشرده شده با الگوریتم LZ4 [3]

هنگام شروع عملیات فشرده‌سازی هر فایل به چند بخش (LZ4 Stream) تقسیم می‌شود که هر کدام از این بخش‌ها شامل عددی جادویی نشان‌دهنده‌ی نحوه و نسخه فشرده‌سازی، مفسر بخش شامل اطلاعاتی مـورد نیـاز فشـرده‌سـازی و همچنین تایید صحت اطلاعات، بلوک‌های داده، مشخص‌کننده پایان بخش و مشخص‌کننده صحت و اتمام صحیح عملیـات. همچنین در این مقاله نویسنده پیشنهادی برای اندازه هر Stream ارائه‌نموده که در آن طول هر کدام را 8مگابایـت اعـلام کرده‌است. البته شایان ذکر است که طبق توضیحات مقاله حداکثر طول ممکن 4 مگابایت می‌باشد کـه نویسـنده پیشـنهاد داده با استفاده از بیت های خالی (درصورت امکان) و تعمیم آن به طول هر Stream، طول حداکثر افزایش یابد.

## 4. AES

این الگوریتم که یکی از سریع‌ترین و امن‌ترین الگوریتم‌های رمزنگاری کلیـد متقـارن مـی‌باشـد در سـال 2001 طـی مقاله‌ای [1] برنده عنوان استاندارد رمزنگاری پیشرفته از سوی موسسه ملی فناوری و استاندارد هـای ایـالات متحـده شـد. پس از کسب این عنوان بخش های امنیتی مختلفی برای رمزکردن داده‌های خود از این الگوریتم استفاده کـرده‌انـد و هـم‌اکنون نیز در ارتباطات مهم از جمله ارتباطات اینترتی نیز از همین الگوریتم استفاده می‌شود. این روش با پیروی از معماری فیستلی معمولا با کلیدهایی با اندازه 128بیتی و بلوک‌هایی به اندازه 128بیت عملیات رمزنگاری را انجام می‌دهد.

## 4.1. سرعت

الگوریتم های کلیدمتقارن همیشه جزء سریع‌ترین الگوریتم‌های رمزنگاری بوده‌اند. در این میان الگوریتم‌های متفاوتی طی سالها ارائه شده اند که هر کدام کاستی‌های مخصوص خود را داشته‌اند. درکنار امنیت بالا که بلوک‌های 128بیتی در آن نقش مهمی را ایفا می‌کنند سرعت الگوریتم نیز مدنظر قرار گرفته‌است. باتوجه به اینکه هر چه اندازه بلوک‌ها بزرگ‌تر باشد سرعت عملیات رمزنگاری نیز افزایش می یابد، در این الگوریتم تعادلی میان امنیت مناسب و سرعت بالا برقرار شده‌است. باوجود امنیت بالا که در بخش بعدی شرح داده خواهدشد سرعت این الگوریتم نیز تا حدی بالاست که بدون اختلال محسوسی در مبادلات درون شبکه‌های بزرگ مانند اینترنت نیز از آن استفاده می‌شود.

### 4.2. امنیت

در سال ۲۰۰۳ سازمان امنیت ملی ایالات متحده AES را به عنوان الگوریتمی مناسب برای رمزنگاری داده‌های طبقه بندی شده اعلام نمود. بهترین حمله گزارش شده تاکنون، در سال ۲۰۱۱ منتشرشد [۴] که تنها ۴ مرتبه از حمله کورکورانه بهتر بود که نشان از امنیت بالای این الگوریتم رمزنگاری دارد.

### 5. ایده‌ی اصلی

پس از بررسی های صورت گرفته بر روی دو الگوریتم مذکور و نیاز به ترکیب این دو برای رسیدن به سرعت، کاهش حجم و امنیت بالای داده ها مناسبت فراوانی میان این دو روش جهت موازی سازی آنها دیده شد. نکته مهمی که بسیار حائز اهمیت است قابلیت هر دو روش در انتقال داده به صورت بلوکی، بدون نیاز به اطلاعات بلوک‌های دیگر است که عملیات موازی‌سازی را بسیار بهینه‌تر می‌نماید.

### 5.1. فشرده سازی و رمزگشایی

با توجه به نیاز به هر چه بهینه‌تر و سریع‌تر بودن، فشرده‌سازی قبل از رمزنگاری قرار می‌گیرد. زیرا در غیراینصورت فشرده سازی برای داشتن نحوه ترکیب بایت‌ها در متن اصلی و انتخاب روش مناسب و همچنین پرکردن Header نیاز به داشتن بخش بزرگی از اوایل متن اصلی را دارد. بنابراین فشرده‌سازی قبل از رمزنگاری انجام می‌شود و متن اصلی برای عملیات فشرده‌سازی به الگوریتم LZ4 تحویل داده می‌شود. الگوریتم LZ4 شروع به فشرده‌سازی داده‌ها در اندازه‌های ۸مگابایتی می‌کند و قبل از اتمام این کار اولین بلوک داده‌ای ۱۲۸بیتی حاصل از نتیجه فشرده‌سازی را به الگوریتم AES جهت رمزنگاری تحویل می‌دهد. درنتیجه اختلاف میان زمان شروع فشرده سازی و رمزنگاری برابر با اختلاف زمان تعیین نحوه فشرده‌سازی و فشرده‌سازی ۱۲۸ بیت اولیه می‌باشد که زمان ناچیزی به حساب می‌آید. به دلیل سریع‌تر بودن عملیات فشرده‌سازی نسبت به عملیات رمزنگاری نیاز به لیستی به عنوان صف ورودی رمزنگاری است تا داده‌هایی را که از عملیات فشرده‌سازی خارج شده و هنوز وارد عملیات رمزنگاری نشده‌اند، در خود نگه دارد. البته این وضعیت در شرایط تک‌نخی رخ خواهد داد. در شرایط چندنخی با اجرای چند عملیات رمزنگاری روی بلوک‌های داده به صورت همزمان، خروجی حاصل از عملیات فشرده‌سازی، مدیریت شده و سرعت پردازش داده‌ها نیز به طور محسوسی افزایش می‌یابد. باتوجه به ویژگی خاص الگوریتم LZ4 در زمان فشرده‌سازی داده‌ها که بسیار سریع است، با اجرای عملیات رمزنگاری به روش چندنخی سرعت کل عملیات رمزنگاری و فشرده‌سازی تقریبا برابر سرعت عملیات فشرده‌سازی با الگوریتم LZ4 (چیزی در حدود ۴۰۰MB/s) است که این نیز با استفاده از روش چندنخی در Stream های الگوریتم LZ4 می‌تواند تا حد بسیار زیادی کاهش یابد. البته باید گفت که اجرای چنین عملیاتی با تعداد زیادی از نخ‌ها نیازمند پردازنده‌های بسیار قوی‌تر می‌باشد.

### 2.5. رمزگشایی و بازسازی

سرعت اجرای پردازش در عمل بازسازی فایل فشرده شده در الگوریتم LZ4 به قدری زیاد است که با تکنولوژی امروزی نیازی به ترکیب و همزمان‌ساختن آن با الگوریتم رمزنگاری دیده نمی‌شود. اما به دلیل امکان توسعه سیستم‌های کامپیوتری در آینده این روش را به عملیات بازسازی تعمیم داده می‌شود.

باتوجه مبتدی بودن عملیات فشرده‌سازی هنگام شروع، در هنگام بازسازی، رمزنگاری مقدم بر عملیات بازسازی فایل فشرده خواهدبود. در اینصورت داده‌های رمزنگاری شده به صورت بلوکی به الگوریتم رمزگشایی ارائه خواهد شد. پس از رمزگشایی اولین بلوک ۱۲۸بیتی داده، الگوریتم بازسازی LZ4 شروع می‌شود. اما در صورتی که عملیات به صورت سیستم تک نخی اجرا شود، ممکن است الگوریتم بازسازی تنها با ۱۲۸ بیت اولیه قادر به شروع عملیات نخواهد بود. زیرا برای شروع عملیات بازسازی داده‌ها، الگوریتم بازسازی نیاز به داشتن تمام اطلاعات Header فایل فشرده را دارد و با توجه به توضیحات ارائه‌شده در بخش شرح الگوریتم LZ4، Header فایل ممکن است هنوز تکمیل نشده‌باشد. بنابراین با توجه ساختار الگوریتم LZ4 امکان شروع عملیات بازسازی داده ها وجود نداشته و باید منتظر دومین بلوک رمزگشایی‌شده بود. حال در صورتی که عملیات رمزگشایی با روش چندنخی اجرا شود هردو بلوک اولیه رمزگشایی‌شده می توانند در یک لحظه و به صورت همزمان نتیجه را به الگوریتم بازسازی ارائه دهند و در اینصورت الگوریتم بازسازی قادر به شروع عملیات خواهد بود. بدیهی است که استفاده از روش چندنخی در سایر قسمت‌ها همانند اصول بیان شده در بخش قبل (فشرده‌سازی و رمزنگاری) موجبات تسریع عملیات را فراهم خواهدساخت.

### 6. نتیجه

در آغاز انتظار از نتیجه مقاله بسیار کمتر از آنچه در نهایت رخداد می‌نمود. انتظار می‌رفت با پشت سرهم قراردادن دو الگوریتم و شروع رمزنگاری تنها زمانی که الگوریتم فشرده سازی قصد نوشتن آخرین نتایج فشرده‌سازی را داشت، سرعت ترکیب فشرده‌سازی و رمزنگاری داده‌ها را بیشتر کرد. اما با مطالعه دقیق‌تر بر روی هر دو الگوریتم نتیجه فراتر از انتظار می‌نمود. زیرا هر دو الگوریتم طوری طراحی شده‌بودند که امکان موازی‌سازی این دو با همدیگر بسیار مناسب و ممکن بود. با ارتباطی که با ارائه‌دهنده الگوریتم LZ4 برقرار شد و اطلاعاتی که ایشان ارائه نمودند نتیجه موازی سازی دو الگوریتم هر چه بهتر نمود. پس از پیاده سازی هر دو الگوریتم و تست آنها بر روی زبان برنامه‌نویسی #C و J AVA و سپس تست نتیجه با ترکیب این دو به روش ارائه شده در مقاله نمایانگر بهینه‌شدن محسوس سرعت عملیات بود.

سطح کاربرد این روش در فناوری اطلاعات به خصوص در سال‌های اخیر بسیار گسترده است. برای مثال استفاده از این روش در ارسال و دریافت نامه‌های الکترونیکی، پشتیبان گیری از اطلاعات پایگاه‌های داده، پشتیبان گیری از اطلاعات سرورهای اطلاعاتی، ذخیره اطلاعات نرم افزارها مانند بازی‌های رایانه‌ای و انتقال اطلاعات حافظه‌های جانبی می‌تواند تاثیر بسزایی در سرعت و امنیت انتقال داده‌ها داشته باشد.

البته این مقاله تنها قصد مطرح‌نمودن گام‌های ابتدایی برای بررسی ترکیب الگوریتم‌های رمزنگاری و فشرده‌سازی و نحوه اجرایی عملیات در الگوریتم‌هایی بدین‌سان را داشت. با بررسی دقیق‌تر و گسترده‌تر در الگوریتم‌های متنوع فشرده-سازی و رمزنگاری می‌توان نتیجه این روش خاص موازی‌سازی را هر چه مناسب‌تر و دقیق‌تر نمود.

### مراجع